\documentclass{jpsj3}
\usepackage{txfonts}

\title{Tunnel Valley Current Filter in the Partially Overlapped Graphene under the Vertical Electric Field}
\author{Ryo Tamura}
\inst{Faculty of Engineering, Shizuoka University, 3-5-1 Johoku, Hamamatsu 432-8561, Japan}

\abst{The tunnel current (TC) and valley current (VC) are crucial in realizing high-speed and energy-saving in next-generation devices. This paper presents the TC and VC link in the partially overlapped graphene. Under the vertical electric field, the two graphene layers have the opposite AB sublattice symmetry, followed by a block on the intravalley transmission. In the allowed intervalley transmission, the difference in the phase of the decay factor prefers only one of the valleys in the output according to the overlapped length.
These results suggest that the band gap with no edge state is a new platform of valleytronics.}

\begin{document}
\maketitle

The tunnel current (TC) and valley current (VC) are clues to satisfy the demand for high-speed and energy-saving devices.
The magnetic tunnel junction serves many applications in the random access memory and hard disk drive \cite{GMR}.
The pure VC unaccompanied by a charge current is expected to be dissipationless \cite{valley-review,valley-review2}.
However, the TC and VC linkage has yet to be a primary target in these fields.
The term 'valley' is a synonym of the inequivalent valence band minimums well separated in the reciprocal space.
Graphene \cite{(1),graphene-review,graphene-review2} and transition metal dichalcogenide (TMD) \cite{TMD-review} are typical valleytronics materials that possess two inequivalent corner points,
$K_+$ and $K_-$, in the Brillouin zone.
A critical issue is the control of the VC, $J_+-J_-$,
where $J_+$ and $J_-$ denote the contributions of valleys $K_+$ and $K_-$ , respectively, to the charge current $J_++J_-$.
A basic unit is a junction with the transmission rate $T_{\nu',\nu}$
from the $K_\nu$ to $K_{\nu'}$ valley.
Based on the  Landauer-B\"{u}ttiker formula (LBF),
$T_\pm= \sum_{\nu'} T_{\pm,\nu'}$
is contribution of the $K_\pm$ valley to the conductance
$(2e^2/h)(T_++T_-)$.
When $ T_\nu \gg T_{-\nu}$,
the junction works as the VC filter (VCF)
that generates the $K_\nu$-polarized VC from the valley-unpolarized current.
The VCF emerges from the line defect \cite{27,28,29,30}, the strain field \cite{31,32,33,34,35,41,42}, zigzag edge states \cite{36,38}, and gate voltage \cite{44,45}.
When $T_{+,-}+T_{-,+} \gg T_{+,+}+T_{-,-}$, the junction indicates the VC reversal (VCR).
The VCR is expected to appear in the zigzag ribbons \cite{61}, $\sqrt{3} \times \sqrt{3}$ superlattice \cite{62,63,64}, and partially
overlapped graphene (po-G) \cite{65, arXiv}.

\begin{figure}
\begin{center}
\includegraphics[width=\linewidth]{2023-Oct-Fig1-2.eps}
\caption{(a)Side view of the partially overlapped graphene (po-G).
(b) Site labels, $j,j_y,A,B,\downarrow$, and $\uparrow$
in the case of $N=5$.
(c),(d),(f)Wavy arrows represent the probability flow.
Refer to the main text.}
\end{center}
\end{figure}

The circularly polarized
light \cite{(9),(10),(11),(12)} and the gate voltage \cite{(7)} lift the valley degeneracy in the TMD junction under the magnetic proximity effect, followed by intervalley difference in the energy band.
When the bulk $K_\nu$ band overlaps the $K_{-\nu}$ band gap at the Fermi level,
the band gap blocks the $K_{-\nu}$ current, and the $K_\nu$-polarized VCF emerges.
In the proper sense, this is {\it not} a tunnel junction because the extended $K_\nu$ states carry the VC.
In contrast, the TMD-based VCF in Ref. \cite{(13)} is an actual tunnel junction
because only the evanescent states mediate the VC in the band gap.
In contrast to the intrinsic gap of the TMD, a nonzero band gap requires
 the vertical electric fields in the bilayer \cite{18,19,22,106,107,112} and substrate effects in the monolayer \cite{23,24,25}.
This may be why the TC-VCF has never been discussed in the graphene system.
This paper proposes the TC-VCF in the po-G shown in Fig. 1(a).
There are many works in the LBF conductance of po-G \cite{67,68,69,70,72,73,74,75,76}, but the VC is
discussed only in Refs. \cite{65,arXiv} .
The top $V_{\rm t}$ and bottom $V_{\rm b}$ gate electrodes exert the vertical electric field and induce the band gap in the overlapped region.
Unlike the TMD junction in Ref. \cite{(13)}, this TC-VCF does not require the magnetic field.
Under the setup of source $V_{\rm L}$ and drain $V_{\rm R}$ electrodes,
the tunnel electrons inevitably pass along the interlayer paths.
Side-contacted armchair nanotubes (sc-ANT) are similar
to the po-G, where the intertube difference in the doping strength
corresponds to the vertical electric field.
The VCF and VCR simultaneously occur at the gap center of the sc-ANT \cite{tamura-2021}.
Although the VCR of the po-G was analyzed for the outside of the energy gap \cite{65,arXiv}, the TC-VCF of the po-G remains unsettled.
When the Fermi level is in the band gap, the edge states \cite{90} and
the whole valence band \cite{87} were theoretically predicted to carry
the VC. This paper shows that the TC is the third possibility.

\begin{figure}
\begin{center}
\includegraphics[width=\linewidth]{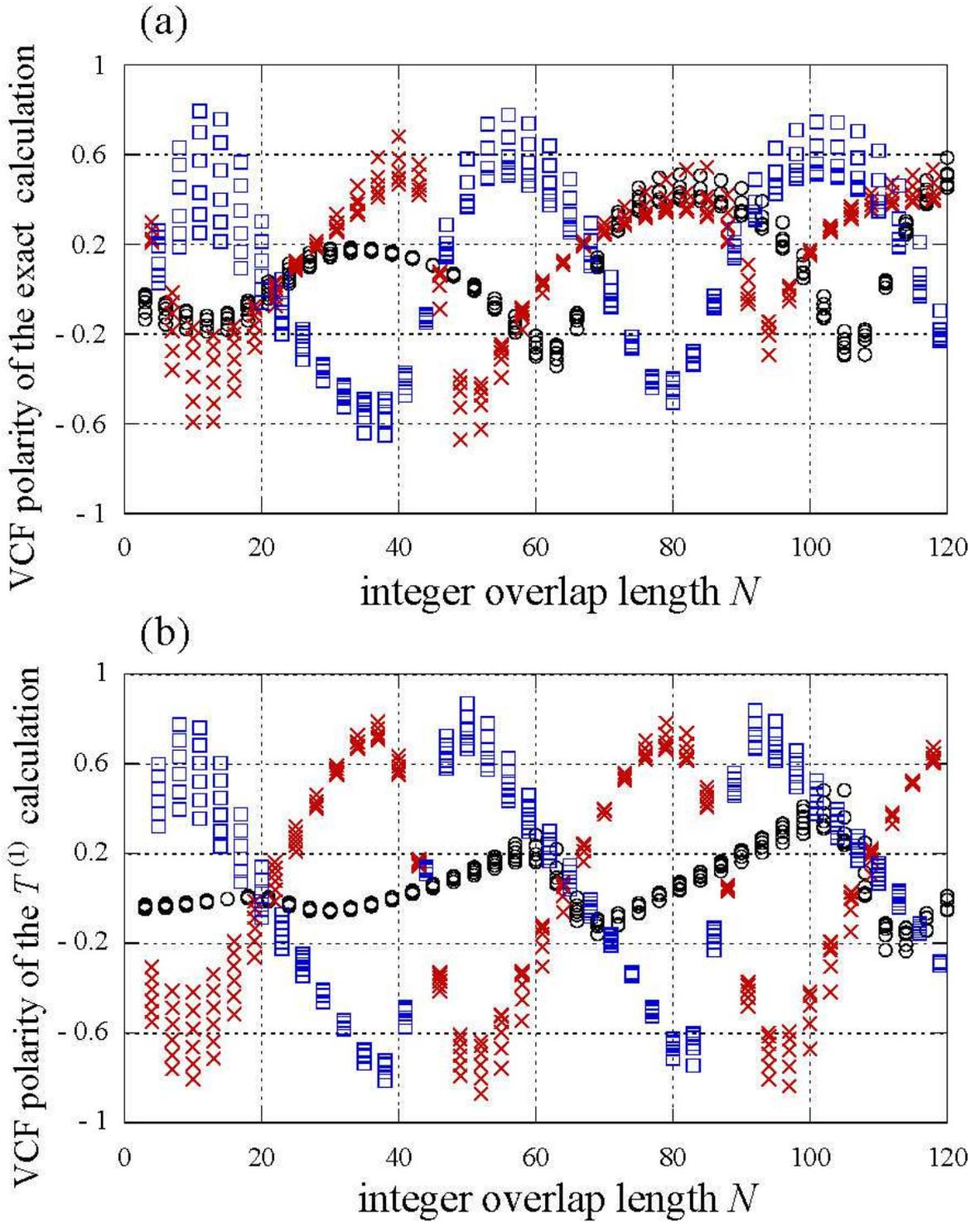}
\caption{
VCF polarity in the case of $\varepsilon=0.35$ eV, and
$E=\pm 0.02, \pm 0.06, \pm 0.1$ eV.
Circles, x marks and squares correspond to the data of mod$(N)=0, 1$,
and 2, respectively,
where mod$(N)$ denotes the remainder of $N$ divided by three.
(a) The exact VCF polarity
$\langle T_+ -T_- \rangle/\langle T_{\rm all} \rangle $
in the case of $N_y=1000$.
(b) The approximate VCF polarity
$(T^{(1)}_+-T_-^{(1)})/T^{(1)}_{\rm all}$.
}
\end{center}
\end{figure}

Figure 1(b) illustrates the atomic structures of the po-G.
The dotted and solid lines
depict the lower ($\downarrow$) and upper ($\uparrow$)
layers, respectively.
Integer indexes ($j$, $j_y$)
and sublattice indexes $(A,B)$ specify the atomic coordinate $(x,y)$ as $x=\frac{a}{2}j$,
$y_{\rm A,\downarrow}=y_{\rm A,\uparrow}=3 a_{\rm c} [j_y+ \frac{1+(-1)^j}{4}]$, $y_{\rm B,\downarrow}=y_{\rm A}-a_{\rm c}$
and $y_{\rm B,\uparrow}=y_{\rm A}+a_{\rm c}$
with the lattice constant $a$
and the bond length $a_{\rm c}=a/\sqrt{3}$.
$(N-2)a/2$ denotes the geometrical overlap length with an integer $N$.
The bilayer region is limited to $1\leq j \leq N-1$.
According to Ref. \cite{TB-parameter}, the TB parameters are $\gamma_0=-3.12$ eV , $\gamma_1=0.377$ eV, $\gamma_3=0.29$ eV, $\gamma_4=0.12$ eV with the standard notation \cite{note-gamma}.
The interlayer site energy difference $2\varepsilon$ represents
the vertical electric field; the site energies are $-\varepsilon$
and $+\varepsilon$ in the $\downarrow$ and $\uparrow$ layers, respectively.
In this paper, we choose $\varepsilon =0.35$ eV,
which opens the band gap in the energy region$|E|< 0.17$ eV.
The comparable band gap was induced in the dual gate experiment \cite{22}.
Applying the exact method of Ref. \cite{tamura-2019} with the periodic boundary condition for the $y$ direction, we can calculate $T_{\nu',\nu}(k_y)$ that denotes the transmission rate with a transverse wave number $k_y$.
The LBF conductance is $(2e^2/h)(2M+1) \langle T_{\rm all} \rangle$ with the $k_y$ average
\begin{equation}
\langle \diamondsuit \rangle
=
\frac{1}{2M+1}\sum_{m=-M}^M \diamondsuit(m \Delta k_y )
\label{k-average}
\end{equation}
and the valley index sum
\begin{equation}
T_{\rm all} =T_++T_-,\;\;
T_\pm =T_{\pm,+}+T_{\pm,-},
\label{valley-sum}
\end{equation}
where $\Delta k_y =\frac{2\pi}{3N_ya_{\rm c}}$,
and the integer $N_y$ stands for the transverse width
$3N_ya_{\rm c}$. In the exact calculation of this paper, $N_y=1000$.
The integer $M$ denotes the maximum effective $k_y$
in the unit of $\Delta k_y$.
Because of the monolayer dispersion relation,
$\sin^2(3k_ya_{\rm c}/2) \leq (E \pm \varepsilon)^2/\gamma_0^2$, and $M$ is close to $N_y||E|-|\varepsilon||/|\pi \gamma_0 |$.
Figure 2(a) shows
the exact VCF polarity $\langle T_+ -T_- \rangle/\langle T_{\rm all} \rangle $ as a function of $N$ for the six energies,
$E=\pm 0.02, \pm 0.06, \pm 0.1$ eV.
All the six energies are in the band gap.
The data are classified according to the remainder of $N$ divided by three, denoted by mod$(N)$.
Circles, x marks, and squares correspond to the data of mod$(N)=0, 1$,
and 2, respectively.
As $E$ approaches the gap center ($E=0$), the decay factor of the wave function decreases, whereas the channel number $2M+1$ increases.
However, the small energy dependence in Fig. 2(a) suggests that these two effects are irrelevant to the ratio $\langle T_\nu \rangle /\langle T_{\rm all} \rangle$.

The effective $k_y$ range is narrow as $ |k_y| < \Delta k_y M \simeq 2||E|-|\varepsilon||/(3a_{\rm c} |\gamma_0|)$.
According to Ref. \cite{tamura-2019}, the effects of $\gamma_3$ and $ \gamma_4$ are minor.
Based on these observations, we calculate
the transmission rate $T^{(1)}_{\nu',\nu}$
under the condition $k_y=0, \gamma_3=\gamma_4=0$,
where the other parameters are the same as the exact calculation.
The $T^{(1)}$ calculation is the analytic continuation
of the analytic formula in Ref. \cite{arXiv}.
In the $T^{(1)}$ calculation,
the wave function in the bilayer region is represented by
\begin{equation}
\left(
\begin{array}{c}
\vec{c}_{\downarrow,j}
\\
\vec{c}_{\uparrow,j}
\end{array}
\right)
=
\sum_{\sigma=\pm}
\sum_{l=\pm}
\sum_{p=\pm}
\lambda_{\sigma,l}^{pj}
\eta^{(p)}_{\sigma,l}
\left(
\begin{array}{c}
\vec{d}^{\;\downarrow}_{\sigma,l}
\\
\vec{d}^{\;\uparrow}_{\sigma,l}
\end{array}
\right),
\label{wf-D}
\end{equation}
where $\eta$ denotes the mode amplitude.
The vector
\begin{equation}
\left[
\;^t\vec{d}_{\sigma,l}^{\;\downarrow},
\;^t\vec{d}_{\sigma,l}^{\;\uparrow}
\right]
=\left[ (\sigma l\alpha_{l}
,\;1),\;\beta_{l}
\left (\sigma l\alpha_{l} \frac{E-\varepsilon}{E+\varepsilon},\;1
\right)
\right],
\label{def-d}
\end{equation}
is the $(\sigma,l)$ mode wave function at $j=0$ sites,
where
\begin{equation}
\alpha_l
=\frac{E+\varepsilon}
{\sqrt{P+lQ}},\;
\beta_{l}=\frac{2\varepsilon E - lQ
}{\gamma_1 (E-\varepsilon)},
\label{def-alpha}
\end{equation}
and
\begin{equation}
P=E^2+\varepsilon^2,\; Q= i
\sqrt{\gamma_1^2\varepsilon^2-(\gamma_1^2+4\varepsilon^2)E^2}.
\end{equation}
In Eq. (\ref{wf-D}),
we use notation $\;^t\vec{c}_{\xi,j}= (A_{\xi,j,0}, B_{\xi,j,0})$. $A_{\xi,j,j_y}$ and $B_{\xi,j,j_y}$
stand for the wave functions at sublattices A and B
with layer index $\xi= \downarrow, \uparrow$.
As $k_y=0$ in the $T^{(1)}$ calculation, the wave functions are independent of $j_y$.
The decay factor $\lambda_{\sigma,l}$ is
\begin{equation}
\lambda_{\sigma,l}
=\mu_{\sigma,l}+\sqrt{\mu_{\sigma,l }^2-1},
\label{def-lambda}
\end{equation}
where $|\lambda_{\sigma,l}| < 1$, $\lambda_{-\sigma,-l}=\lambda_{\sigma,l}^*$, and
\begin{equation}
\mu_{\sigma,l}
=\frac{-1}{2}\left(1+ \frac{l\sigma}{|\gamma_0|} \sqrt{P+lQ} \right).
\label{def-mu}
\end{equation}
As $\lambda_{\sigma,l} \simeq |\lambda_{\sigma,l}| e^{\sigma i 2\pi/3}$,
$\sigma$ represents the valley index in the bilayer region \cite{note2}.
When we fix the valley $\sigma$ and decay direction $p$,
there remain two degenerate evanescent modes that are complex conjugate to each other.
The index $l$ corresponds to this degeneracy, where $\alpha_-=\alpha_+^*$, and $\beta_-=\beta_+^*$.
In the left monolayer region $(j \leq 0)$, the wave function
is approximated by
\begin{equation}
\vec{c}_{\downarrow,j}^{\;(0)}
=
\sum_{\nu=\pm}
\sum_{p=\pm}
e^{ip\nu\frac{2}{3}\pi j}
\eta^{(p)}_{\nu,{\rm L}}
\left(
\begin{array}{c}
\nu
\\
1
\end{array}
\right).
\label{wf-mu}
\end{equation}
The scattering matrix at the boundary $j=0$
is
defined by
\begin{equation}
\left(
\begin{array}{c}
\vec{\eta}^{\;(+)} \\
\vec{\eta}^{\;(-)}_{\rm L}
\end{array}
\right)
=
\left(
\begin{array}{cc}
r_\downarrow, & t_\downarrow \\
\tilde{t}_\downarrow, & \tilde{r}_\downarrow
\end{array}
\right)
\left(
\begin{array}{c}
\vec{\eta}^{\;(-)} \\
\vec{\eta}^{\;(+)}_{\rm L}
\end{array}
\right),
\label{sl-exact}
\end{equation}
where
$\;^t\vec{\eta}^{\;(p)}=(\eta_{+,+}^{(p)},\;\eta_{-,+}^{(p)},\;\eta_{+,-}^{(p)},\;\eta_{-,-}^{(p)})
$, and $\;^t\vec{\eta}^{\;(p)}_{\rm L}
=(\eta_{+,{\rm L}}^{(p)},\;\eta_{-,{\rm L}}^{(p)})
$.
Appling Eqs. (\ref{wf-D}) and (\ref{wf-mu})
to the boundary conditions $\vec{c}_{\downarrow,0}=\vec{c}_{\downarrow,0}^{\;(0)}$, $\vec{c}_{\uparrow,0}=0$, and
$\vec{c}_{\downarrow,1} =\vec{c}_{\downarrow,1}^{\;(0)}$
with the approximation
$\lambda_{\sigma,l} \simeq e^{i \sigma \frac{2}{3} \pi}$,
we obtain
\begin{eqnarray}
\left(
\begin{array}{cc}
r_\downarrow, & t_\downarrow \\
\tilde{t}_\downarrow, & \tilde{r}_\downarrow
\end{array}
\right)
&=&
\frac{2}{c_B}
\left(
\begin{array}{ccc}
\frac{\alpha_+^2}{v_+}&
\frac{-\alpha_+\alpha_-}{v_+}&
\frac{\alpha_+}{v_+} \\
\frac{-\alpha_-\alpha_+}{v_-}&
\frac{\alpha_-^2}{v_-}&
\frac{-\alpha_-}{v_-} \\
\alpha_+&
-\alpha_-&
1
\end{array}
\right)
\otimes u_B
\nonumber \\
&&+
\frac{2}{c_A}
\left(
\begin{array}{ccc}
\frac{1}{v_+}&
\frac{1}{v_+}&
\frac{1}{v_+} \\
\frac{1}{v_-}&
\frac{1}{v_-}&
\frac{1}{v_-} \\
1
&
1 &
1
\end{array}
\right)
\otimes
u_A
-{\bf 1}_6,
\label{sl}
\end{eqnarray}
where $\bf{1}_n$ is the $n$ dimensional unit matrix, $\otimes$ stands for the Kronecker product,
\begin{equation}
v_l=
\frac{\alpha_l}{\beta_{(-l)}}
\left(
\beta_{-}-\beta_+
\right),
\label{def-v}
\end{equation}

\begin{equation}
u_{\;_A^B}
=
\; \frac{1}{2} \left(
\begin{array}{cc}
1 & \pm 1
\\
\pm 1 & 1
\end{array}
\right),\;
c_{\;_A^B}=
1+\frac{\alpha_+^{\pm 1} \beta_-+\alpha_-^{\pm 1} \beta_+}{\beta_--\beta_+}.
\label{u-c}
\end{equation}
Indexes $B$ and $A$ in Eqs. (\ref{sl}) correspond to the periodic sublattice localization discussed in Ref. \cite{arXiv} .
The probability flow is determined by $\eta_{\sigma,l}, \lambda_{\sigma,l}$, $\eta_{\sigma,l}^{(p)}$, and $v_l$ \cite{note}.
The 6 $\times$ 6 matrix of Eq. (\ref{sl-exact}) is not unitary,
but $\tilde{r}_\downarrow^*\tilde{r}_\downarrow =\bf{1}_2$,
corresponding to the perfect reflection in the case of infinite $N$.
We also obtain an approximate formula
\begin{equation}
\left(
\begin{array}{cc}
r_\uparrow, & t_\uparrow \\
\tilde{t}_\uparrow, & \tilde{r}_\uparrow
\end{array}
\right)
= \frac{1}{V}\otimes{\bf 1}_2
\left(
\begin{array}{cc}
r^{\;\prime}_\downarrow, & t^{\;\prime}_\downarrow \\
\tilde{t}^{\;\prime}_\downarrow, & \tilde{r}^{\;\prime}_\downarrow
\end{array}
\right)
V\otimes{\bf 1}_2,
\label{sr}
\end{equation}
for the scattering at the boundary $j=N$,
where $V$ is the 3 $\times$ 3 diagonal matrix
with the elements $V_{1,1}=\beta_1$, $V_{2,2}=\beta_2$,
and $V_{3,3}=1$.
We transform
$r_\downarrow,t_\downarrow,\tilde{r}_\downarrow,\tilde{t}_\downarrow,
$ into
$r^{\;\prime}_\downarrow,t^{\;\prime}_\downarrow,\tilde{r}^{\;\prime}_\downarrow,\tilde{t}^{\;\prime}_\downarrow$ ,
by replacing $(\alpha_l,\beta_l)$
with $(\alpha_l^{\;\prime},\beta_l^{\;\prime})=\left(\frac{E-\varepsilon}{E+\varepsilon}\alpha_l, \frac{1}{\beta_l}\right)$.
Combining Eqs. (\ref{sl}) and (\ref{sr}),
we derive the approximate transmission rate $T^{(1)}_{\nu',\nu}=|t^{(1)}_{\nu',\nu}|^2$ where
\begin{equation}
t^{(1)}= \tilde{t}_{\uparrow}\Lambda^N
\left({\bf 1}_4-r_{\downarrow}\Lambda^N r_{\uparrow}\Lambda^N\right)^{-1}
t_{\downarrow}
\label{sl-sr}
\end{equation}
and $\Lambda$ is the diagonal matrix
with the elements defined by Eq. (\ref{def-lambda}).

\begin{figure}
\begin{center}
\includegraphics[width=\linewidth]{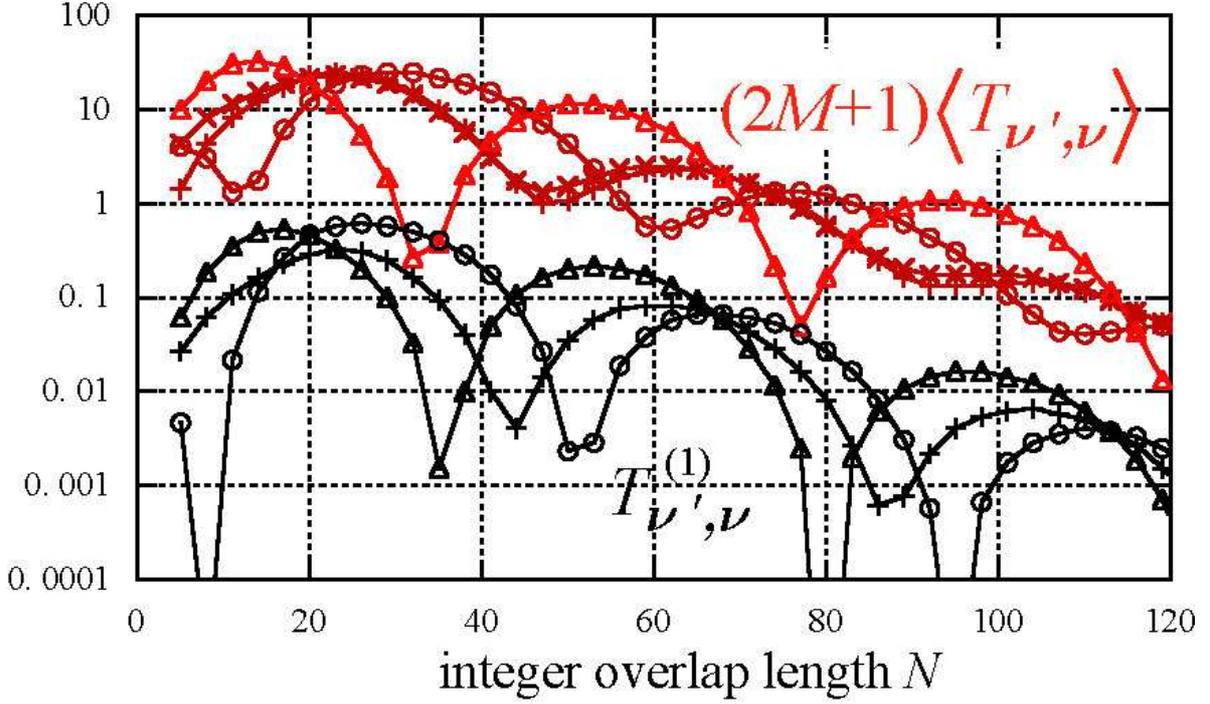}
\caption{
Exact VC components $(2M+1) \langle T_{\nu',\nu} \rangle$ (red symbols)
and the approximate transmission rate $T^{(1)}_{\nu',\nu}$
(black symbols) in the case of mod$(N)=$2, $\varepsilon=0.35$ eV, $E=0$, and $N_y=1000$. Under these conditions, $M=35$.
The triangles, circles, + and x marks represent
the cases $(\nu',\nu)=(+,-),(-,+),(+,+)$, and $(-,-)$,respectively,
where $\nu'$ and $\nu$ correspond to the valleys of the output and input flows, respectively.
Since $T^{(1)}_{+,+}=T^{(1)}_{-,-}$ holds
at the zero $E$, we omit the black x marks.
}
\end{center}
\end{figure}

Figure 2(b) shows
the approximate VCF polarity
$(T^{(1)}_+-T_-^{(1)})/T^{(1)}_{\rm all}$.
The decaying factor $\lambda_{\sigma,l}$ is not
a real number, and thus causes the periodic oscillation, reproducing the oscillation of
the Fig. 2(a) as follows.
The $K_+$ ($K_-$) peak appears when $N\simeq 37$, and 79
( $N\simeq 10, 52$, and 97) in the case of mod$(N)=1$.
Compared with this case, the VCF polarity is opposite and small in the case of
 mod$(N)=$2 and 1, respectively.
Figure 3 displays the valley-resolved components of the square data in Fig. 2.
Red and black symbols in Fig. 3 represent the $(2M+1) \langle T_{\nu',\nu} \rangle$ and $T^{(1)}_{\nu',\nu}$, respectively,
under the conditions of mod$(N)=2$, $E=0$, $\varepsilon= 0.35$ eV, and $N_y=1000$.
The $E$, $\varepsilon$ and $N_y$ determine that $M=35$.
Since the energy dependence is small in Fig. 2, we mainly discuss the zero $E$ .
The triangles, circles, + and x marks represent
the cases $(\nu',\nu)=(+,-),(-,+),(+,+)$, and $(-,-)$,respectively.
Since $T^{(1)}_{+,+}=T^{(1)}_{-,-}$ at the zero $E$,
we omit the black x marks.
Owing to the narrow effective $k_y$ range, $T^{(1)}_{\nu',\nu}$ reproduces the essential characteristics of $\langle T_{\nu',\nu} \rangle$ as in the relation between Figs. 2(a) and 2(b).
Considering the periodic boundary condition, we neglect the edge mode \cite{36,38,61}. This assumption is reasonable under the following conditions.
The number of the right-going zigzag edge modes is only one per valley in the gap center \cite{89}.
In the case of $N < 60$, the edge mode has little influence on the VGR polarity in Fig. 2 because the corresponding values of $(2M+1)|\langle T_{+} - T_{-} \rangle|$ are much larger than one in Fig. 3.
Interestingly, the VCF and VCR simultaneously appear in Fig. 3.
The $K_\nu$-polarized VCF occurs when $T_{\nu,-\nu} > T_{-\nu,\nu}$.
Contrarily, the intravalley transmission is irrelevant to the VCF as $T_{+,+} \simeq T_{-,-}$.
Notably, the relation $\langle T_{\nu',\nu} \rangle = \langle T_{\nu,\nu'} \rangle$ does {\it not} hold.
Here, we fix the notation $\langle T_{\nu',\nu}(E,\varepsilon) \rangle $
according to the three rules;
(1) The left and right subindexes represent the valleys of the transmitted and incident waves, respectively.
(2) The direction of the $x$ axis is the same as that of the incident wave.
(3) The layer with the incident (transmitted) wave
is labeled by $\downarrow$ ($\uparrow$ ) and
has the site energy $-\varepsilon$ ($+\varepsilon$).
Figure 1(c) represents the incident and transmitted waves corresponding to $\langle T_{\nu',\nu}(E,\widetilde{\varepsilon}) \rangle$.
The complex conjugate operation (CCO) of the wave function
transforms Fig. 1(c) into Fig. 1(d) and Fig. 1(e),
where the rules (2) and (3) are not applied to Fig. 1(d)
for the comparison.
First, The CCO causes reversal both in the probability flow
and in the valley.
This results in the change of the subindex from $(\nu',\nu)$ to
$(-\nu,-\nu')$
according to the rule (1).
Second, the rule (2) causes
the inversion of the $x$ axis,
accampanied by the change $(-\nu,-\nu') \rightarrow (\nu,\nu')$.
Lastly, the rule (3) changes the assignment of $\downarrow$ and $\uparrow$,
and thus $\varepsilon =-\widetilde{\varepsilon}$ in Fig. 1(e).
Comparing Fig. 1(c) with Fig. 1(e), we prove
\begin{equation}
\langle T_{\nu',\nu}(E,\varepsilon) \rangle
=\langle T_{\nu,\nu'}(E,-\varepsilon) \rangle
\label{T-sym-1}
\end{equation}
and find that $\langle T_{\nu,\nu'} \rangle \neq
\langle T_{\nu',\nu} \rangle$ with a nonzero $\varepsilon$.
With the transformation
$(A'_{\downarrow},B'_\downarrow)=(A^*_{\downarrow},-B^*_\downarrow)$ and $(A'_{\uparrow},B'_\uparrow)=(-A^*_{\uparrow},B^*_\uparrow)$, we can also prove
\begin{equation}
T_{\nu',\nu}^{(1)}(E,\varepsilon)
= T_{-\nu',-\nu}^{(1)}(-E,-\varepsilon).
\label{T-sym-2}
\end{equation}
Because of Eqs. (\ref{T-sym-1}) and (\ref{T-sym-2}),
$T_{+,+} \simeq T_{-,-}$ holds when $E \simeq 0$.

When $N$ is large and $E$ is close to zero,
$T^{(1)}_{\nu',\nu}$ is approximated by $T^{(2)}_{\nu',\nu}=|t^{(2)}_{\nu',\nu}|^2$, where
\begin{eqnarray}
t^{(2)}_{\nu',\nu} &=& (\tilde{t}_{\uparrow}\Lambda^N t_{\downarrow})_{\nu',\nu} |_{E=0} \nonumber \\
&= &
\frac{1}{i} {\rm Re} \left[ \left(
\frac{\alpha_0}{|\tilde{c}_B|^2}
-
\frac{\nu\nu'}{\alpha_0|\tilde{c}_A|^2}
\right) (\tilde{\lambda}_+^N+\tilde{\lambda}_-^N)
\right] \nonumber \\
&&
+ \frac{1}{i} \left(
\frac{\nu}{\tilde{c}_B^*\tilde{c}_A}
-
\frac{\nu'}{\tilde{c}_A^*\tilde{c}_B}
\right){\rm Re}[ \tilde{\lambda}_+^N-\tilde{\lambda}_-^N
],
\label{t2}
\end{eqnarray}
$\alpha_0$, $\tilde{c}_B$, $\tilde{c_A}$, and
$\tilde{\lambda}_\pm$ stand for $\alpha_{+,+}$, $c_B$, $c_A$,
and $\lambda_{\pm,+}$, respectively, in the case of zero $E$;
$\alpha_0=\varepsilon/\sqrt{\varepsilon^2+i|\varepsilon|\gamma_1}$,
$\tilde{c}_B=1+(\alpha_0-\alpha_0^*)/2$,
$\tilde{c}_A=1+(\alpha_0^*-\alpha_0)/(2|\alpha_0|^2)$,
$\tilde{\lambda}_\pm=\mu_\pm+\sqrt{\mu_\pm^2-1}$, and
$\mu_\pm=-\frac{1}{2}\mp\frac{\varepsilon}{2|\gamma_0|\alpha_0}$.
For an intuitive picture of Eq. (\ref{t2}), we define the wave function
matching $I_{\nu',\nu}$ as
\begin{equation}
I_{\nu',\nu} \equiv \frac{1}{2i}\sum_{\sigma,l}
\beta_l^{-1}\lambda_{\sigma,l}^N
(\nu',1)\vec{d}_{\sigma,l}^{\;\uparrow}
(\nu,1)
\vec{d}_{\sigma,l}^{\;\downarrow}.
\label{I}
\end{equation}
The factors
$(\nu',1)\vec{d}_{\sigma,l}^{\;\uparrow}
\lambda_{\sigma,l}^N
$
and
$(\nu,1)\vec{d}_{\sigma,l}^{\;\downarrow}
$
correspond
to the transmission at $j=0$ and $j=N$, respectively.
We choose the coefficient $1/(2i\beta_l)$ to obtain the relation
\begin{eqnarray}
\lim_{\frac{\varepsilon}{\gamma_1}
\rightarrow \pm \infty}
t^{(2)}_{\nu',\nu}
&=&
\lim
_{\frac{\varepsilon}{\gamma_1}
\rightarrow \pm \infty}
I_{\nu',\nu} |_{E=0 }
\label{t-I}
\\
&=&
\frac{4}{i}
\delta_{\nu',-\nu}
{\rm Re}\left (
\tilde{\lambda}^N_{\nu\frac{\varepsilon}{|\varepsilon|}}
\right ).
\label{t3}
\end{eqnarray}
In Fig. 4, $T^{(1)}_{-\nu,\nu}$ data in Fig. 3
are compared with the $T^{(2)}_{-\nu,\nu}$
and $16|{\rm Re}(\tilde{\lambda}_\pm^N)|^2$.
$T^{(2)}$ reproduces $T^{(1)}$, especially when $N > 30$.
Although the $\varepsilon$ is slightly smaller
than the $\gamma_1$,
the factor $|{\rm Re}(\tilde{\lambda}_\pm^N)|^2$
qualitatively explains the oscillation,
contrasting with the monotonic decay caused by the
factor $|\tilde{\lambda}|^{2N}$.
When the sign of $\varepsilon$ is reversed, the VCF polarity
is also reversed, similar to other gated graphene systems \cite{28,30,36,38}.
In the case of Eq. (\ref{t-I}),
the factor $(E-\varepsilon)/(E+\varepsilon)=-1$ in $ (\nu', 1)\vec{d}_{\sigma,l}^{\;\uparrow }$
blocks the intravalley $(\nu'=\nu$) transmission.
Additionally, the factors
$ (\nu, 1)\vec{d}_{\sigma,l}^{\;\downarrow *}$
and $ (-\nu, 1)\vec{d}_{\sigma,l}^{\;\uparrow }$,
which correspond to the VCR $(\nu'= -\nu$),
simultaneously grow
in Eq. (\ref{t-I})
under the condition
of $\nu\sigma l \varepsilon/|\varepsilon| = 1$,
leading us to Eq. (\ref{t3}).

\begin{figure}
\begin{center}
\includegraphics[width=\linewidth]{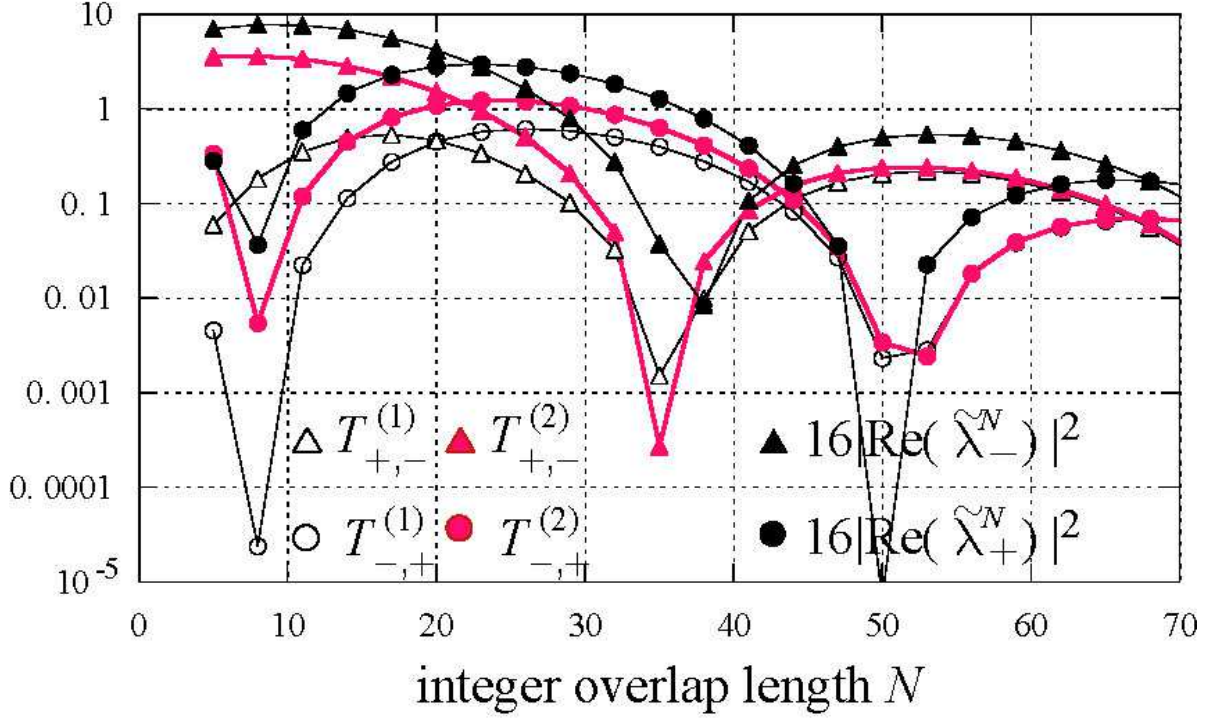}
\caption{The approximate intervalley transmission
rates, $T^{(1)}_{\pm,\mp}$, $T^{(2)}_{\pm,\mp}$,
and $16|{\rm Re}(\tilde{\lambda}_\mp^N)|^2$, in the
case of $\varepsilon=0.35$ eV, $E=0$, and mod$(N)=2$.}
\end{center}
\end{figure}

The TC-VCF and TC-VCR have been ignored in the graphene system.
Given the successful application of tunneling magnetoresistance, the TC deserves to be surveyed \cite{GMR}.
Moreover, we can measure the tunneling time \cite{tunnel-1,tunnel-2} in the band gap using optical measurement.
The pump polarized pulse light is shed on the left monolayer, creating a high population in the $K_\nu$ valley \cite{50,52,53,57,58,Dixit}.
This high population drives the $K_\nu$ electrons to diffuse to the valley-unpopulated area, i.e., to the right monolayer.
The probe pulse light illuminates the right monolater region and induces the second harmonic generation (SHG) \cite{Dixit,51}.
The SHG signal of the $K_{-\nu}$ valley corresponds to the VCR and thus is a signature of the TC.
The delay time of the probe light gives us the time scale information.
The absolute value of the decay factor is close to one and permits a significant variation in the tunnel barrier thickness.
Under the condition $(2M+1) \langle T_{\pm,\mp} \rangle \gg 1$ in Fig. 3, the thickness reaches about 15 nm,
indicating that the po-G is a new platform for exploring the tunneling process.


\begin{thebibliography}{99}
\bibitem{note-gamma}
The sign of $\gamma_0$ and
$\gamma_{4}$ are reversed compared to Ref. \cite{TB-parameter} by the transformation $(A',B')=(A,-B)$.
\bibitem{note2}
In this paper, the square root
of a complex number $X=|X|e^{i\theta}$
is defined by
$\sqrt{X}=\sqrt{|X|}e^{i\frac{\theta}{2}}$
with the phase range $-\pi < \theta \leq \pi$.
The real numbers $Y$ and $Z$
in the formula $Y+ilZ=\frac{1}{|\gamma_0|}\sqrt{P+lQ}$
become positive under this range of $\theta$.
Since $|E/\gamma_0| \ll 1$ and $|\varepsilon/\gamma_0| \ll 1$,
$Y$ and $Z$ are much less than one.
Accordingly, $\mu_{\sigma,l}^2-1 \simeq -\frac{3}{4}+\frac{l\sigma}{2}Y+\frac{i}{2}\sigma Z$,
$\sqrt{\mu_{\sigma,l}^2-1} \simeq i\frac{\sqrt{3}}{2}\left(\sigma -\frac{l}{3}Y \right)$, $\lambda_{\sigma,l} \simeq \exp(i\sigma 2\pi/3)-\frac{l\sigma}{2}
Y-\frac{i}{2}\left( \sigma Z+ \frac{l}{\sqrt{3}}Y \right)$,
and $|\lambda_{\sigma,l}|^2 \simeq 1-\frac{\sqrt{3}}{2}Z < 1$.
\bibitem{note}
In the $T^{(1)}$ calculation,
the probability flow is represented by
$J={\rm Im}[ \;^t\vec{c}_{\downarrow, j}^{\;*} h \vec{c}_{\downarrow, j-1}
+\;^t\vec{c}_{\uparrow, j}^{\;*} h \vec{c}_{\uparrow, j-1}]$.
Here, $h$ denotes the Hamiltonian matrix connecting $\vec{c}_{j}$ with $\vec{c}_{j-1}$.
The  elements are $h_{1,1}=h_{2,2}=0$, and $h_{1,2}=h_{2,1}=\gamma_0$.
In the $T^{(1)}$ calculation, $\gamma_3=\gamma_4=0$, and $\gamma_1$ does not
 connect  $\vec{c}_{j}$ with $\vec{c}_{j-1}$.
Using Eq. (\ref{wf-D}), we can prove
$J =2\gamma_0\sum_{\sigma,l,p} {\rm Im}[\sigma v_l\lambda_{\sigma,l}^{-1}\eta_{\sigma,l}^{(p)}
\eta_{-\sigma,-l}^{(-p)*}]
$.
\end{thebibliography}
\end{document}